\documentclass[pra,twocolumn,showpacs,preprintnumbers,amsmath,amssymb]{revtex4}

\usepackage{graphicx}
\usepackage{dcolumn}
\usepackage{bm}
\usepackage{color}
\usepackage{float}
\begin{document}
\preprint{APS/123-QED}
\noindent \textbf{Comment on "Reentrant Localization Transition in a Quasiperiodic Chain"}

In their Letter~\cite{RO21}, the authors found an interesting reentrant localization phenomenon in a one-dimensional dimerized
lattice with quasiperiodic disorder, i.e., the system undergoes a second localization transition at a higher disorder strength. They claimed that this reentrant feature would appear in the case of staggered disorder, but not for the uniform ones. In this Comment, we show that such reentrant feature can also appear for the uniform disorder when larger system parameters are taken into account.

The model has two sublattices  A and B and the intra- and intercell hopping strengths are denoted by $t_1$ and $t_2$. The dimerization strength is represented by $\delta=t_2/t_1$ and on-site quasiperiodic disorder strength is $\lambda_A$ ($\lambda_B$). In the case of uniform disorder, $\lambda_A=\lambda_B=\lambda$. By using the standard approach  followed by the Letter's authors\cite{RO21}, we analyze the inverse participation ratio (IPR) and the normalized participation ratio (NPR)~\cite{EV08,LI17,LI20} defined as $IPR_n=\sum^L_i{|\phi^i_n|^4}$ and $NPR_n=(L\sum^L_i{|\phi^i_n|^4})^{-1}$ to identify the localization transition. Here, $n$ is the eigenstate index, $\phi^i_n$ is the amplitude of eigenstate on $i$th site and $L$ is the system size.

\begin{figure}[!htbp]
(a)\includegraphics[width=1.5in]{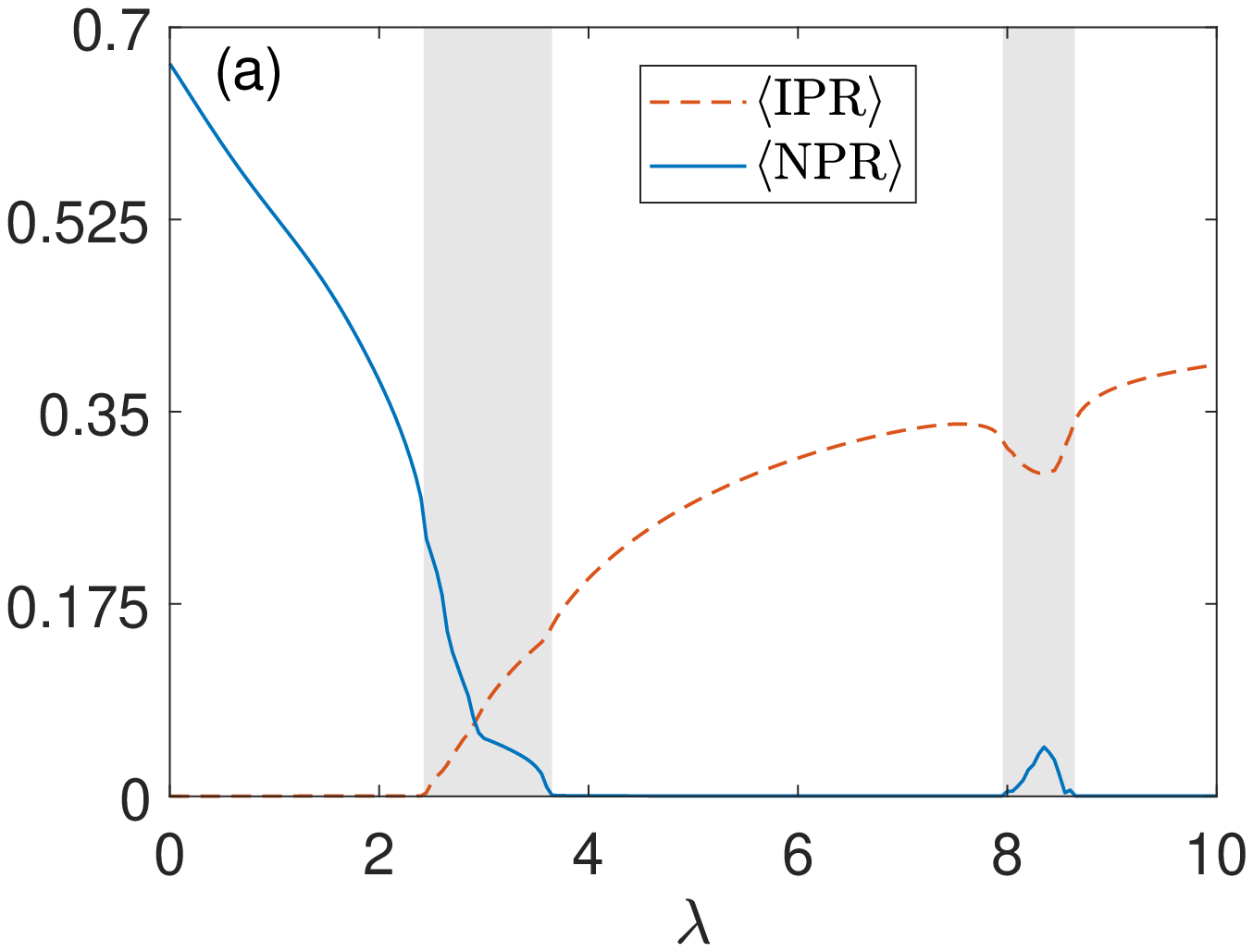}
(b)\includegraphics[width=1.5in]{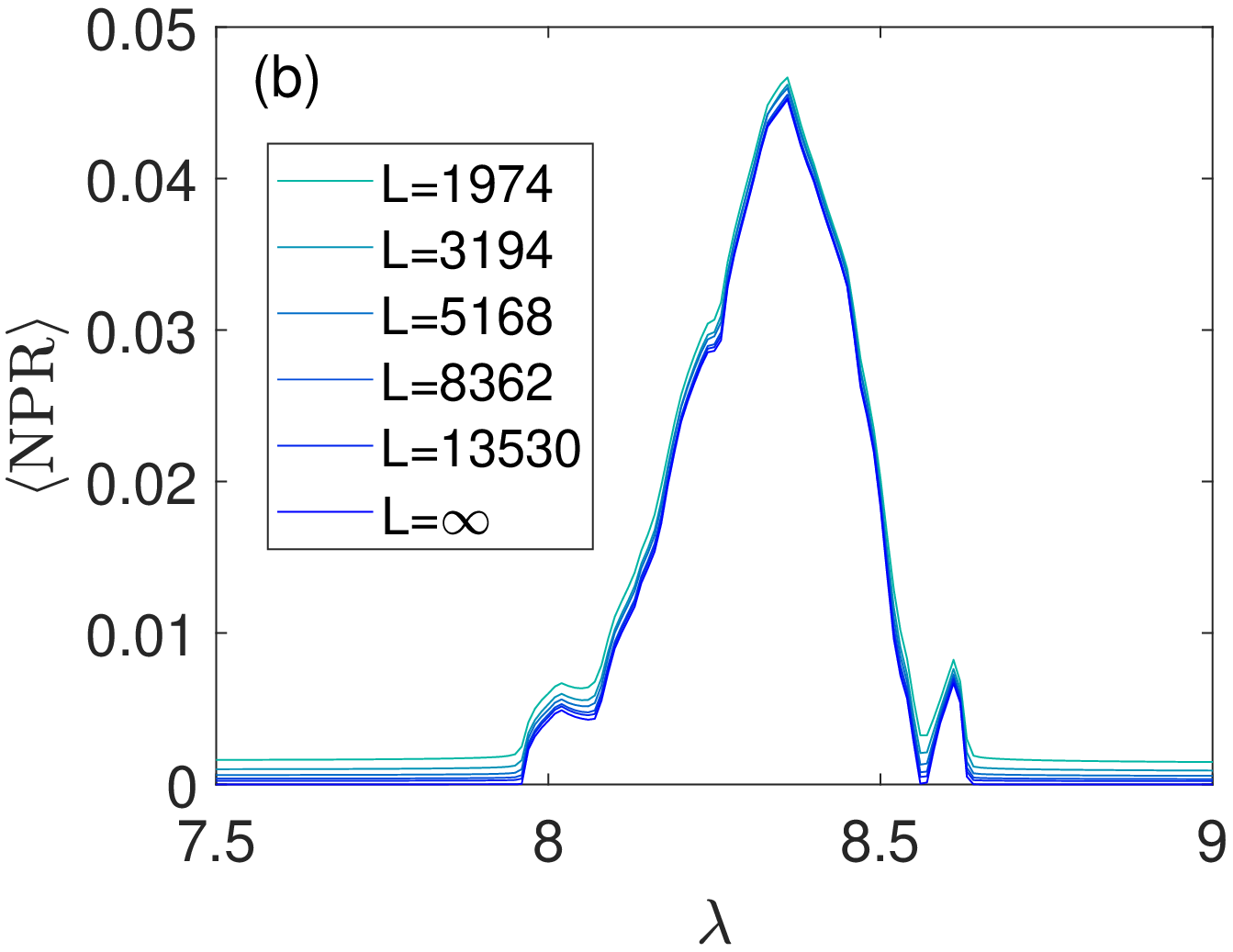}
\caption{(a) The spectrum averaged $\rm \langle{IPR}\rangle$ and $\rm \langle{NPR}\rangle$ for $\delta=25$ and $L=13530$. The shaded regions indicate the
critical or the intermediate regimes. (b) The $\rm \langle{NPR}\rangle$ at $\delta=25$ for $L=1974, 3194, 5168, 8362, 13530$,
and $\infty$ (light to deep blue).}\label{Fig1}
\end{figure}

Without lost of generality, we take $\delta=25$ and $L=13530$ as an example. The corresponding spectrum averaged $\rm \langle{IPR}\rangle$ and $\rm \langle{NPR}\rangle$ are plotted in Fig.\ref{Fig1}(a). It shows there are two shaded regions ($2.43<\lambda<3.64$ and $7.96<\lambda<8.64$), where both the $\rm \langle{IPR}\rangle$ and the $\rm \langle{NPR}\rangle$ are finite. This signifies such regions are  critical or intermediate~\cite{RO21}. When $3.64<\lambda<7.96$ and $\lambda>8.64$, the $\rm \langle{IPR}\rangle$ is finite and $\rm \langle{NPR}\rangle$ approaches to zero, which indicates all states are localized. Therefore, there happens a reentrant localization phenomenon as $\lambda$ changes from relative small value to relative large ones. Near the second shaded region, $\rm \langle{NPR}\rangle$ for several system sizes are plotted in Fig.\ref{Fig1}(b). The values of $\rm \langle{NPR}\rangle$ at $L\to\infty$ are obtained with the method of finite size extrapolation~\cite{RO21}. It shows the second critical region is stability.

\begin{figure}[!htbp]
\includegraphics[width=2.0in]{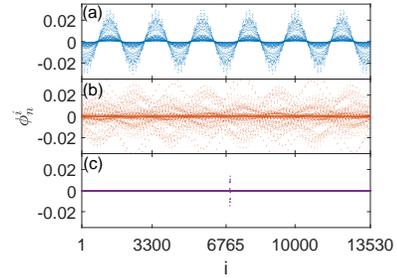}
\caption{Three typical eigenstates for $\lambda=8.35$, $\delta=25$ and $L=13530$, where the eigenstate index is (a)$n=13$, (b)$n=940$ and (c) $n=1607$, respectively.
}\label{Fig2}
\end{figure}

To visually illustrate state localization properties in the second critical regime, three typical states are plotted in Fig.\ref{Fig2} for $\delta=25$, $\lambda=8.35$ and $L=13530$. States in Figs.\ref{Fig2}(a) and (b) are extended or critical states, where they spread over the whole system. The state in Fig.\ref{Fig2}(c) is a localized state, where it stays in few of sites. For them, $\rm \langle{NPR}\rangle$ are equal to $0.220276, 0.071638$ and $1.464536\times10^{-4}$, respectively. Fig.\ref{Fig2} shows extended, critical and localized states coexist, so the parameter space ($\lambda=8.35$, $\delta=25$) belongs to critical regimes.

\begin{figure}[!htbp]
(a)\includegraphics[width=2.1in]{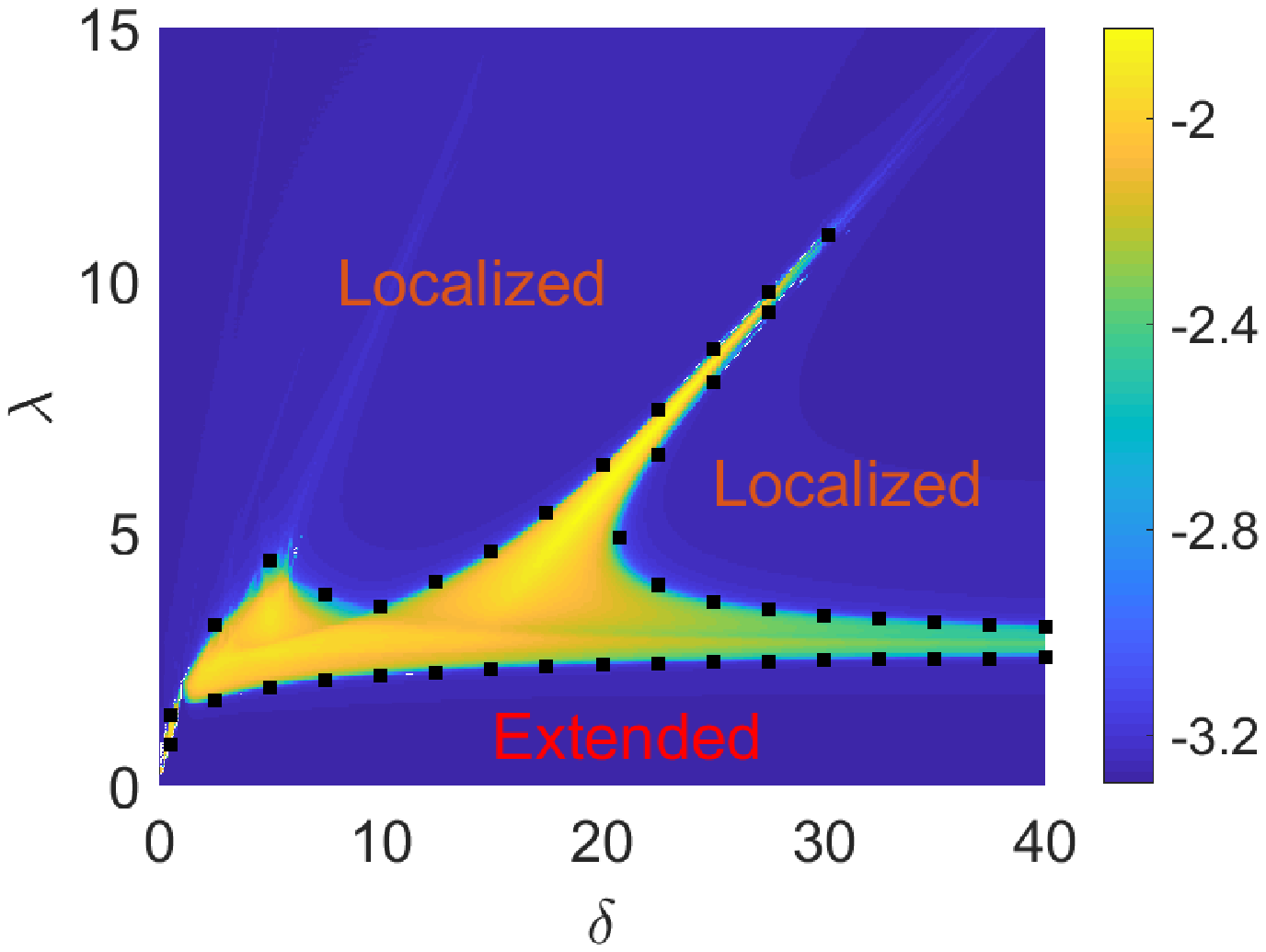}
(b)\includegraphics[width=2.0in]{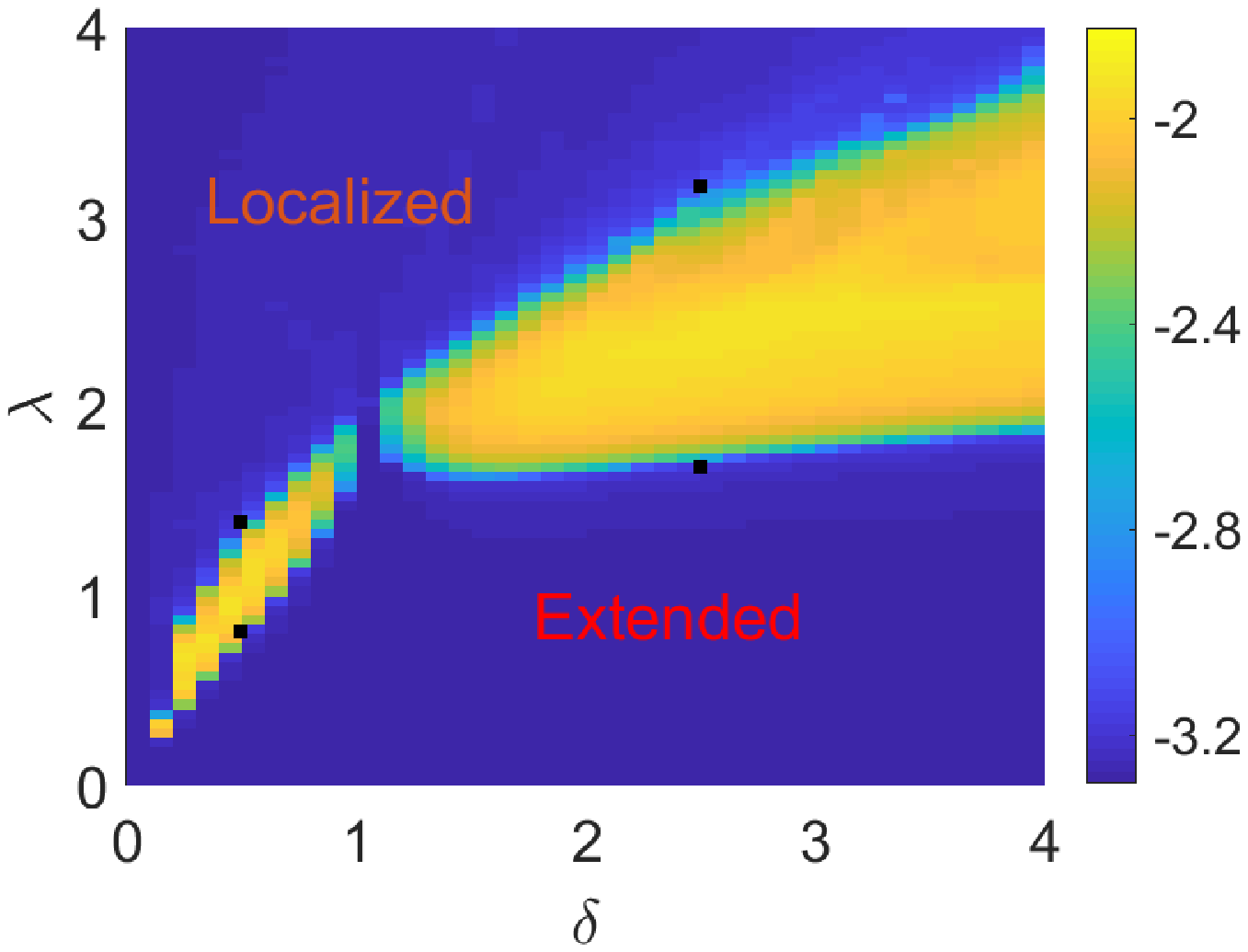}
\caption{(a)The phase diagrams in $\delta$ and $\lambda$ plane. (b) Partial enlarger of (a) as $\delta\in[0,4]$. The filled black squares
are the data points obtained by examining the $\rm \langle{IPR}\rangle$ and the $\rm \langle{NPR}\rangle$ in the thermodynamic limit. The colorbar indicates the values
of $\eta$ at $L=1974$.}\label{Fig3}
\end{figure}

The quantity $\eta=\log_{10}[\langle{IPR}\rangle\times\langle{NPR}\rangle]$
is used to distinguish different phase regions~\cite{RO21,LI20}.
The phase diagram is given in Fig.\ref{Fig3}. There are the critical region (yellow region bounded by
filled black squares), the fully extended and the fully localized regions (blue regions). The boundaries
(filled squares) of the critical region is located by examining the
values of the $\rm \langle{IPR}\rangle$ and the $\rm \langle{NPR}\rangle$ in the thermodynamic limit~\cite{RO21}.
Fig.3(a) shows that for intermediate values of $\delta$, i.e., $20.8<\delta<30.2$,  the system
undergoes two localization transitions through two critical phases as the quasiperiodic disorder strength $\lambda$ increases. At relative larger dimerization ($\delta>30.2$), the system may change to ''isolated dimers or double-wells'', all corresponding states are localized in space and there are no second localization transitions. Therefore, in intermediate values of $\delta$ ($20.8<\delta<30.2$), the competition between dimerization and quasiperiodic disorder leads to the reentrant localization transition. Fig.\ref{Fig3}(b) is the partial enlarger of (a) as $\delta\in[0,4]$, which is just that studied in Ref.~\cite{RO21}. As only the small range of
$\delta$ is considered, the reentrant localization phenomenon is omitted in Ref.~\cite{RO21}.

Summary, we show that the reentrant localization phenomenon can happen even for the uniform disorder.

\vspace{0.5cm}
\noindent Longyan Gong$^{1,2}$, Hui Lu$^{1}$ and Weiwen Cheng$^{3}$\\
$^{1}$College of Science, Nanjing University of Posts and Telecommunications, Nanjing, 210003, China\\
$^{2}$New Energy Technology Engineering of Jiangsu Province, Nanjing University of Posts and Telecommunications, Nanjing, 210003, China\\
$^{3}$Institute of Signal Processing and Transmission, Nanjing University of Posts and Telecommunication
Nanjing 210003, China\\
\end{document}